\documentclass[11pt]{article}

\usepackage[margin=1in]{geometry}
\usepackage{setspace}
\usepackage{amsmath, amssymb, amsthm}
\usepackage{mathtools}
\usepackage{xurl}
\usepackage[font=small,labelfont=bf]{caption}
\usepackage{graphicx} 
\usepackage{tikz}
\usepackage{pgfplots}
\pgfplotsset{compat=1.18}
\usetikzlibrary{shapes.geometric, arrows.meta, positioning, calc}
\usetikzlibrary{patterns,decorations.pathreplacing,calligraphy}
\usepackage{booktabs}
\usepackage{float}
\usepackage[numbers]{natbib}
\usepackage{hyperref}
\hypersetup{
    colorlinks=true,
    linkcolor=blue,
    citecolor=blue,
    urlcolor=blue
}

\usepackage{algorithm}
\usepackage{algorithmic}
\usepackage{titlesec}

\begin{document}

\begin{center}
    {\LARGE \textbf{Second Thoughts: How 1-second subslots transform CEX-DEX Arbitrage on Ethereum}}
    
    \vspace{1.5em}
    
    \begin{tabular}{ccccc}
    {\large Aleksei Adadurov} &  &{\large Sergey Barseghyan} &  &{\large Anton Chtepine} \\
    nuconstruct &  & nuconstruct &  & nuconstruct
    \end{tabular}
    
    \vspace{1em}
    
    \begin{tabular}{ccccc}
    {\large Antero Eloranta} &  & {\large Andrei Sebyakin} &  &{\large Arsenii Valitov}\\
    nuconstruct &  & nuconstruct &  & nuconstruct
    \end{tabular}
    
    \vspace{1em}
\end{center}

\vspace{2em}

\begin{center}
    \textbf{Abstract}
\end{center}

This paper examines the impact of reducing Ethereum slot time on decentralized exchange activity, with a focus on CEX-DEX arbitrage behavior. We develop a trading model where the agent's DEX transaction is not guaranteed to land, and the agent explicitly accounts for this execution risk when deciding whether to pursue arbitrage opportunities.  

We compare agent behavior under Ethereum's default 12-second slot time environment with a faster regime that offers 1-second subslot execution. The simulations, calibrated to Binance and Uniswap v3 data from July to September 2025, show that faster slot times increase arbitrage transaction count by 535\% and trading volume by 203\% on average.

The increase in CEX-DEX arbitrage activity under 1-second subslots is driven by the reduction in variance of both successful and failed trade outcomes, increasing the risk-adjusted returns and making CEX-DEX arbitrage more appealing. \\


\newpage

\section{Introduction}

One of the most active sources of DEX volume on Ethereum is arbitrage between centralized and decentralized exchanges. The strategy is simple: when the prices between a CEX (e.g., Binance) and a DEX (e.g., Uniswap) differ, a trader can exploit such an opportunity by simultaneously buying on the cheaper venue and selling on the more expensive one, thus capturing the spread. However, this strategy is not risk-free: while the CEX leg is executed almost instantaneously, the DEX leg must wait for block inclusion, which is not guaranteed. DEX transactions can fail for multiple reasons, such as insufficient priority fees, another trader capturing the same opportunity, latency, or even the block builder censoring the transaction.  

The risk of executing a DEX leg and its relationship to slot times is the focus of our paper. We model a trading agent who knows that their on-chain transaction may fail to land and who incorporates this uncertainty into their entry and exit decisions. The essence of the setup is as follows: the agent faces a known probability $\alpha$ of successfully executing the DEX leg, and if it fails, they must decide how to manage their resulting delta exposure.

The motivation for the paper comes from the interest in understanding how Ethereum slot time reduction would affect the largest portion of DEX volume. We consider a theoretical protocol that operates in 1-second subslots, providing a faster feedback loop for traders compared to Ethereum's native 12-second slots. The question we address is: how does faster execution change arbitrageurs' behavior, and what are the downstream effects on DEX activity? \\

\noindent \textbf{Related Literature.} \textit{CEX-DEX Arbitrage Empirics.} A growing number of papers study the scale and structure of cross-domain arbitrage between centralized and decentralized exchanges. \cite{cexdex1} provides a systematic analysis of CEX-DEX arbitrage on Ethereum, identifying over 157,000 arbitrage transactions between May and July 2023. Their research shows that nearly all CEX-DEX arbitrages execute at the top of blocks, confirming that these opportunities require priority access. The paper also documents significant market concentration: a single searcher (0xa69) captured 55.7\% of all identified arbitrages, while beaverbuild dominated market share at 41.8\%. Starting from this foundation, \cite{cexdex4} extends the analysis to 19 months of data (from August 2023 to March 2025), estimating the total extracted value of \$233.8 million across 7.2 million CEX-DEX arbitrages. Their findings reveal increasing centralization, with three searchers capturing three-quarters of both volume and extracted value, and demonstrate that searcher profitability is tightly linked to vertical integration with block builders.

The relationship between CEX-DEX arbitrage and block building competition is examined in \cite{cexdex3} and \cite{cexdex5}. \cite{cexdex3} shows that CeFi-DeFi arbitrage opportunities materially alter MEV-Boost auction bid profiles, with builders who have access to CEX-DEX flow gaining decisive advantages during periods of high volatility. \cite{cexdex5} confirms that integrated high-frequency trading builders who extract top-of-block opportunities are favored to win block auctions precisely when the price volatility is increased, contributing to the observed concentration where three builders produce 80\% of all MEV-Boost blocks.

\textit{Block Time and LP Losses}. The theoretical relationship between block time and arbitrage losses is formalized by the Loss-Versus-Rebalancing (LVR) framework, which predicts that LP losses scale with $\sqrt{BT}$, where $BT$ is block time. However, \cite{cexdex2} extends this model to incorporate transaction costs and shows that the $\sqrt{BT}$-approximation breaks down on chains with significant base fees. Their simulations show that when transaction costs are non-trivial, LP losses depend on multiple interacting factors: base fees, swap fees, and block times, with the relative importance varying by pool characteristics.

\textit{Preconfirmations}. Recent work has begun to formalize preconfirmations as a distinct service layer in Ethereum’s execution pipeline. While preconfirmations represent the closest analog to our theoretical protocol currently being researched, they differ fundamentally in scope: preconfirmations guarantee transaction inclusion but do not provide intermediate state commitments during execution. The framework in \cite{preconf1} introduces a fulfillment-delivery paradigm borrowed from supply-chain logistics, in which specialized fulfillers (i.e., block builders) issue real-time commitments that transactions will be included, while deliverers (i.e., validators) guarantee eventual on-chain confirmation. Within this paradigm, preconfirmations are interpreted as a tool for reducing transaction execution risk without requiring validator centralization, because economically specialized entities (e.g., builders) can provide commitments while the underlying validator set remains decentralized and compatible with solo staking. 

Empirical evidence on the economic impact of preconfirmations for validators is still limited. Using early mainnet data from August 2025, \cite{preconf2} measures how running preconfirmation services affects realized block value and finds that, at the 75th percentile of blocks, validators with preconfirmations observe higher block value than standard validators. However, the study emphasizes that the sample is small and the results are not yet statistically significant. 

An alternative line of work studies preconfirmations when validators themselves act as preconfirmation providers instead of relying on external gateway architectures. Under this assumption, \cite{preconf3} shows that preconfirmations create new MEV capture channels and materially alter timing games: searchers face heightened risk of early termination and therefore bid more aggressively earlier in the block auction, which in turn incentivizes earlier transaction submission. The same work argues that validators can preconfirm their own transactions to capture on-chain arbitrage and liquidation opportunities, estimating that a validator with 1\% stake penetration could earn roughly \$345,600 in incremental annual MEV, representing about a 25\% increase—higher than the current increase from timing games that add roughly 10\% to MEV revenue. \ 

Finally, \cite{preconf4} analyzes whether execution preconfirmations can be made economically sustainable for proposers relative to the current MEV-Boost regime. The paper models Dependent Sub-Slot Auctions in which each preconfirmation depends on expected future block and preconfirmation value, and shows that these auctions can increase expected proposer revenue even in the absence of explicit preconfirmation-specific tips. In their design, a builder who preconfirms the $\text{N}^{\text{th}}$ transaction captures all subsequent revenue in the slot, effectively tying preconfirmation incentives to the future flow of MEV. \\ 

\noindent \textbf{Our Contribution.} We complement the literature in two ways. First, unlike the empirical studies that document realized arbitrage outcomes, we model the \textit{decision-making} problem facing a CEX-DEX arbitrageur under execution uncertainty. This perspective reveals why observed arbitrage activity may substantially understate the latent demand for faster execution: many profitable opportunities are foregone not because they do not exist but because rational agents decline to attempt them given current confirmation times. Second, we focus on execution guarantees that can be provided without protocol-level changes, making our analysis applicable to a range of possible mechanism designs, including but not limited to preconfirmations.

Most importantly, we model a realistic trading agent who accounts for execution uncertainty. This agent does not simply attempt every profitable opportunity. Instead, they evaluate the risk-adjusted expected value of entry, considering both the probability of DEX leg failure and the costs of closing an unsuccessful position. This behavioral realism proves crucial for understanding how faster execution affects market outcomes.

Our results show that faster execution substantially increases on-chain arbitrage activity. For a risk-averse agent with 35\% probability of landing the DEX leg, transaction counts increase by 535\% when moving from 12-second to 1-second confirmation times. Volume increases by up to 203\%. \\

\section{DEX Price Interpolation}

The simulation framework models the interaction between centralized and decentralized exchange prices over time. In this paper, we study the difference between confirmation intervals: under Ethereum's default configuration, DEX transactions can be executed every 12 seconds at slot boundaries, while under faster execution regimes, transactions can be executed every second within subslots. Historical DEX prices are readily available and can be used for 12-second slot benchmark simulations. For a faster regime, we need a framework to interpolate these prices for 1-second subslots. 

Our framework includes three components. First, we start with a historical price for a current slot. It is also a price for the initial subslot. Next, we derive a price $p^{\text{DEX}}(t_{i})$ of the $i$-th subslot from a price $p^{\text{DEX}}(t_{i-1})$ of the $(i-1)$-th subslot. To do this, we apply (1) arbitrage transactions from the previous slot if present, (2) CEX-DEX price reversion to model how DEX prices adjust to off-chain information between arbitrage events, and (3) noise trading to capture non-arbitrage DEX activity. Each component addresses a specific modeling challenge and, thus, we run simulations with different combinations to assess robustness.

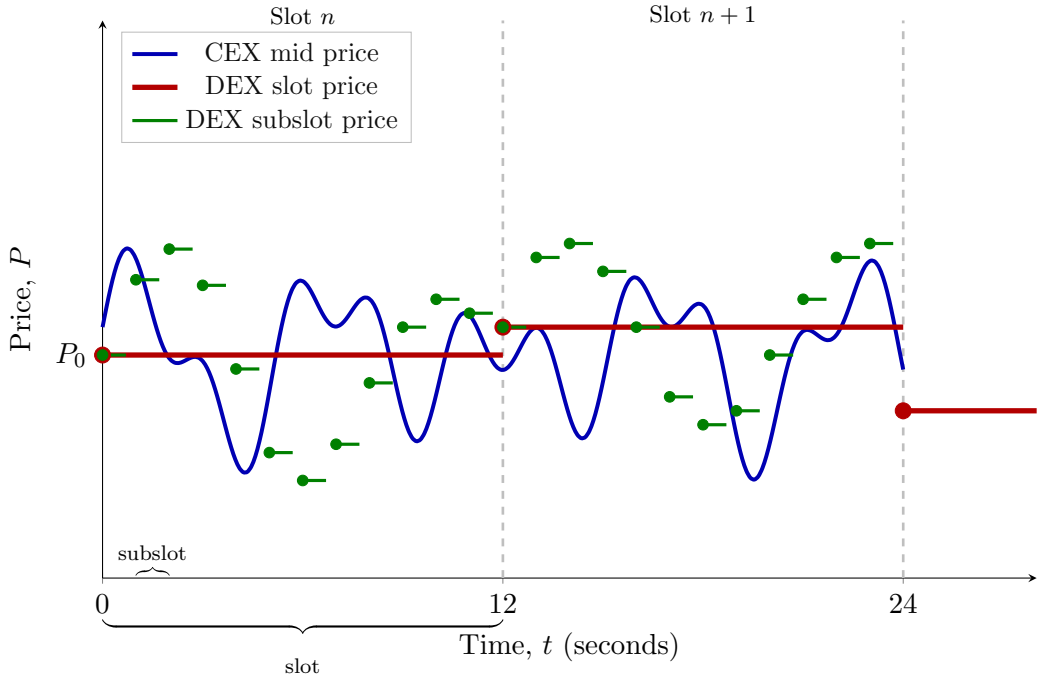
\begin{figure}[htbp]
\centering
\begin{tikzpicture}
\begin{axis}[
    width=14cm,
    height=9cm,
    xlabel={Time, $t$ (seconds)},
    ylabel={Price, $P$},
    xmin=0, xmax=28,
    ymin=0.92, ymax=1.12,
    xtick={0,12,24},
    xticklabels={$0$, $12$, $24$},
    ytick={1.00},
    yticklabels={$P_0$},
    grid=none,
    axis lines=left,
    clip=false,
    legend style={
        at={(0.02,0.98)},
        anchor=north west,
        font=\small,
        fill=white,
        fill opacity=0.95,
        draw=gray!50
    },
]

\addplot[gray!50, dashed, line width=1pt, forget plot] coordinates {(12,0.92) (12,1.12)};
\addplot[gray!50, dashed, line width=1pt, forget plot] coordinates {(24,0.92) (24,1.12)};

\addplot[blue!70!black, line width=1.5pt, smooth, samples=200, domain=0:24] 
    {1.00 + 0.02*sin(deg(x*1.2)) + 0.015*sin(deg(x*2.5)) + 0.01*cos(deg(x*0.8))};
\addlegendentry{CEX mid price}

\addplot[red!70!black, line width=2pt] coordinates {(0,1.00) (12,1.00)};
\addlegendentry{DEX slot price}

\addplot[red!70!black, line width=2pt, forget plot] coordinates {(12,1.01) (24,1.01)};
\addplot[red!70!black, line width=2pt, forget plot] coordinates {(24,0.98) (28,0.98)};

\addplot[only marks, mark=*, mark size=3pt, red!70!black, forget plot] coordinates {(0,1.00) (12,1.01) (24,0.98)};

\addplot[green!50!black, line width=1.2pt] coordinates {(0,1.00) (0.7,1.00)};
\addlegendentry{DEX subslot price}

\addplot[green!50!black, line width=1.2pt, forget plot] coordinates {(1,1.027) (1.7,1.027)};
\addplot[green!50!black, line width=1.2pt, forget plot] coordinates {(2,1.038) (2.7,1.038)};
\addplot[green!50!black, line width=1.2pt, forget plot] coordinates {(3,1.025) (3.7,1.025)};
\addplot[green!50!black, line width=1.2pt, forget plot] coordinates {(4,0.995) (4.7,0.995)};
\addplot[green!50!black, line width=1.2pt, forget plot] coordinates {(5,0.965) (5.7,0.965)};
\addplot[green!50!black, line width=1.2pt, forget plot] coordinates {(6,0.955) (6.7,0.955)};
\addplot[green!50!black, line width=1.2pt, forget plot] coordinates {(7,0.968) (7.7,0.968)};
\addplot[green!50!black, line width=1.2pt, forget plot] coordinates {(8,0.99) (8.7,0.99)};
\addplot[green!50!black, line width=1.2pt, forget plot] coordinates {(9,1.01) (9.7,1.01)};
\addplot[green!50!black, line width=1.2pt, forget plot] coordinates {(10,1.02) (10.7,1.02)};
\addplot[green!50!black, line width=1.2pt, forget plot] coordinates {(11,1.015) (11.7,1.015)};

\addplot[green!50!black, line width=1.2pt, forget plot] coordinates {(12,1.01) (12.7,1.01)};
\addplot[green!50!black, line width=1.2pt, forget plot] coordinates {(13,1.035) (13.7,1.035)};
\addplot[green!50!black, line width=1.2pt, forget plot] coordinates {(14,1.04) (14.7,1.04)};
\addplot[green!50!black, line width=1.2pt, forget plot] coordinates {(15,1.03) (15.7,1.03)};
\addplot[green!50!black, line width=1.2pt, forget plot] coordinates {(16,1.01) (16.7,1.01)};
\addplot[green!50!black, line width=1.2pt, forget plot] coordinates {(17,0.985) (17.7,0.985)};
\addplot[green!50!black, line width=1.2pt, forget plot] coordinates {(18,0.975) (18.7,0.975)};
\addplot[green!50!black, line width=1.2pt, forget plot] coordinates {(19,0.98) (19.7,0.98)};
\addplot[green!50!black, line width=1.2pt, forget plot] coordinates {(20,1.00) (20.7,1.00)};
\addplot[green!50!black, line width=1.2pt, forget plot] coordinates {(21,1.02) (21.7,1.02)};
\addplot[green!50!black, line width=1.2pt, forget plot] coordinates {(22,1.035) (22.7,1.035)};
\addplot[green!50!black, line width=1.2pt, forget plot] coordinates {(23,1.04) (23.7,1.04)};

\addplot[only marks, mark=*, mark size=2pt, green!50!black, forget plot] coordinates {
    (0,1.00) (1,1.027) (2,1.038) (3,1.025) (4,0.995) (5,0.965) (6,0.955) (7,0.968) (8,0.99) (9,1.01) (10,1.02) (11,1.015)
    (12,1.01) (13,1.035) (14,1.04) (15,1.03) (16,1.01) (17,0.985) (18,0.975) (19,0.98) (20,1.00) (21,1.02) (22,1.035) (23,1.04)
};

\node[anchor=south, font=\footnotesize] at (axis cs:6,1.115) {Slot $n$};
\node[anchor=south, font=\footnotesize] at (axis cs:18,1.115) {Slot $n+1$};

\draw[decorate, decoration={calligraphic brace, amplitude=5pt, mirror}, line width=0.8pt] 
    (axis cs:0,0.905) -- (axis cs:12,0.905);
\node[anchor=north, font=\scriptsize] at (axis cs:6,0.895) {slot};

\draw[decorate, decoration={calligraphic brace, amplitude=3pt}, line width=0.6pt] 
    (axis cs:1,0.921) -- (axis cs:2,0.921);
\node[anchor=south, font=\scriptsize] at (axis cs:1.5,0.923) {subslot};

\end{axis}
\end{tikzpicture}
\caption{\centering Price dynamics and subslot structure within Ethereum slots. The CEX mid price (blue) updates continuously, while the DEX slot price (red) updates only at slot boundaries every 12 seconds. DEX subslot prices (green) are interpolated from the CEX price at 1-second intervals within each slot.}
\label{fig:price_subslots}
\end{figure}

The subslot price interpolation mechanism is depicted in \hyperref[fig:price_subslots]{Figure 1}. All other market mechanics remain constant, allowing us to isolate the effect of confirmation speed. 

We use data from July to September 2025, including millisecond-level best bid and ask prices from Binance and all swap transactions from Uniswap v3 ETH-USDC pools at the 30 basis point, 5 basis point, and 1 basis point tiers.

\subsection{Noise Trading}

It is known that arbitrageurs are not the only participants trading on DEXs. Hedgers, retail users, and other non-arbitrage traders generate transaction flow that moves prices independently of CEX-DEX spreads. In order to include other market participants and avoid overstating the predictability of DEX price movements, we incorporate "noise trading".

We classify historical Uniswap v3 transactions as either arbitrage or noise based on their characteristics. A trade is considered arbitrage if it satisfies all of the following four conditions: (1) it is the first transaction in a block, (2) the pre-trade CEX-DEX price discrepancy exceeds the pool fee, (3) the trade moves the DEX price toward the CEX price, and (4) the post-trade price discrepancy remains equal to or above the pool fee. All other transactions are classified as noise.

From July to September 2025 data, we estimate probability distributions for both the number of noise trades per block and the price impact of individual noise trades. Price impacts are rounded to the nearest basis point, and observations with absolute impact exceeding 30 basis points are excluded as outliers. During simulation, we sample a random number of noise trades with random price impacts every 12 seconds and distribute them randomly across the DEX execution subslots.

\subsection{CEX-DEX Price Reversion}
\label{sec:twotwo}

DEX prices can be adjusted toward CEX prices through channels other than direct arbitrage. For example, informed traders may update their limit orders, or market makers may rebalance inventory. In order to capture this gradual price convergence, we implement a regression-based reversion mechanism.

The idea is to estimate how much the DEX price typically moves in response to a given CEX price change and then use this relationship to revert DEX prices to CEX between arbitrage events. Let $p^{\text{CEX}}(t)$ and $p^{\text{DEX}}(t)$ denote the CEX and DEX prices at time $t$. Then, for consecutive timestamps $t_{i-1}$ and $t_i$, define simple returns:

\[r^{\text{CEX}}_i = \frac{p^{\text{CEX}}(t_i) - p^{\text{CEX}}(t_{i-1})}{p^{\text{CEX}}(t_{i-1})}, \quad r^{\text{DEX}}_i = \frac{p^{\text{DEX}}(t_i) - p^{\text{DEX}}(t_{i-1})}{p^{\text{DEX}}(t_{i-1})}\]

\noindent Then, we partition of the time axis into semi-open intervals $F_k = (\tau_{k-1}, \tau_k]$ and estimate a linear model within each interval:

\[r^{\text{DEX}}_i = \beta^{(k)}_0 + \beta^{(k)}_1 r^{\text{CEX}}_i + \varepsilon_i, \quad \varepsilon_i \overset{\text{iid}}{\sim} \mathcal{N}(0, \sigma^2_k)\]

\noindent The OLS estimate of the coefficients is as follows:

\[
\hat{\beta}^{(k)}_1 = \frac{\sum\limits_{(r^{\text{CEX}}, r^{\text{DEX}}) \in F_k} (r^{\text{CEX}} - \bar{r}^{\text{CEX}}_k)(r^{\text{DEX}} - \bar{r}^{\text{DEX}}_k)}{\sum\limits_{(r^{\text{CEX}}, r^{\text{DEX}}) \in F_k} (r^{\text{CEX}} - \bar{r}^{\text{CEX}}_k)^2}, \quad \hat{\beta}^{(k)}_0 = \bar{r}^{\text{DEX}}_k - \hat{\beta}^{(k)}_1 \bar{r}^{\text{CEX}}_k
\]

\noindent For any timestamp $t_i$ within interval $F_k$, where the CEX price is observed but the DEX price is not yet updated, we predict:

\[\hat{r}^{\text{DEX}}_i = \hat{\beta}^{(k)}_0 + \hat{\beta}^{(k)}_1 r^{\text{CEX}}_i\]

\noindent and obtain the reverted DEX price from the most recent observed DEX price $p^{\text{DEX}}(t_{i-1})$:

\[
\hat{p}^{\text{DEX}} (t_i) = p^{\text{DEX}} (t_{i-1}) (1 + \hat{r}^{\text{DEX}}_i)\]

\noindent It is also important to note that the coefficients are recalibrated at each interval boundary to allow the CEX-DEX relationship to vary over time.

\section{Modeling the Trading Agent}

\begin{figure}[htbp]
\centering
\begin{tikzpicture}
\begin{axis}[
    width=14cm,
    height=8cm,
    xlabel={Time (seconds)},
    ylabel={Price (USDC per ETH)},
    xmin=0, xmax=72,
    ymin=3385, ymax=3420,
    xtick={0,12,24,36,48,60,72},
    ytick={3390,3395,3400,3405,3410,3415},
    grid=both,
    grid style={line width=0.2pt, draw=gray!30},
    major grid style={line width=0.4pt, draw=gray!50},
    legend pos=north west,
    legend style={font=\small, fill=white, fill opacity=0.9, draw=gray!50},
]

\addplot[dashed, gray!60, line width=0.5pt, forget plot] coordinates {(12,3385) (12,3420)};
\addplot[dashed, gray!60, line width=0.5pt, forget plot] coordinates {(24,3385) (24,3420)};
\addplot[dashed, gray!60, line width=0.5pt, forget plot] coordinates {(36,3385) (36,3420)};
\addplot[dashed, gray!60, line width=0.5pt, forget plot] coordinates {(48,3385) (48,3420)};
\addplot[dashed, gray!60, line width=0.5pt, forget plot] coordinates {(60,3385) (60,3420)};

\fill[green!25, opacity=0.6] (axis cs:6,3400) rectangle (axis cs:12,3408);
\fill[green!25, opacity=0.6] (axis cs:28,3391) rectangle (axis cs:36,3414);
\fill[orange!25, opacity=0.6] (axis cs:48,3393) rectangle (axis cs:54,3412);
\fill[green!25, opacity=0.6] (axis cs:63,3408) rectangle (axis cs:72,3415);

\addplot[
    blue!70!black, 
    line width=1.3pt,
    smooth
] coordinates {
    (0,3400) (1,3401) (2,3402.5) (3,3401) (4,3399) (5,3397) 
    (6,3398) (7,3400) (8,3403) (9,3405) (10,3407) (11,3408)
    (12,3406) (13,3404) (14,3402) (15,3400) (16,3398) (17,3396)
    (18,3394) (19,3393) (20,3392) (21,3391) (22,3390) (23,3389)
    (24,3390) (25,3392) (26,3394) (27,3396) (28,3398) (29,3401)
    (30,3404) (31,3407) (32,3409) (33,3411) (34,3413) (35,3414)
    (36,3412) (37,3410) (38,3408) (39,3407) (40,3406) (41,3405)
    (42,3404) (43,3403) (44,3402) (45,3401) (46,3400) (47,3399)
    (48,3398) (49,3397) (50,3396) (51,3395) (52,3394) (53,3393)
    (54,3394) (55,3396) (56,3398) (57,3400) (58,3402) (59,3404)
    (60,3406) (61,3408) (62,3410) (63,3411) (64,3412) (65,3413)
    (66,3414) (67,3415) (68,3414) (69,3413) (70,3412) (71,3411) (72,3410)
};
\addlegendentry{CEX mid price}

\addplot[
    red!70!black, 
    line width=1.3pt,
    const plot
] coordinates {
    (0,3400) (12,3400) (12,3406) (24,3406) (24,3391) (36,3391) 
    (36,3412) (48,3412) (48,3400) (60,3400) (60,3408) (72,3408)
};
\addlegendentry{DEX price}

\node[font=\footnotesize, text=green!50!black, align=center] at (axis cs:32,3418) {Arbitrage\\opportunity};
\draw[->, green!50!black, line width=0.8pt] (axis cs:32,3416) -- (axis cs:32,3414.5);

\draw[->, thick, red] (axis cs:23.1,3404) -- (axis cs:23.1,3393);
\node[font=\scriptsize, red, align=center, anchor=west] at (axis cs:16.8,3398) {DEX\\update};

\node[font=\scriptsize, gray] at (axis cs:6,3383) {Slot $n$};
\node[font=\scriptsize, gray] at (axis cs:18,3383) {Slot $n{+}1$};
\node[font=\scriptsize, gray] at (axis cs:30,3383) {Slot $n{+}2$};
\node[font=\scriptsize, gray] at (axis cs:42,3383) {Slot $n{+}3$};
\node[font=\scriptsize, gray] at (axis cs:54,3383) {Slot $n{+}4$};
\node[font=\scriptsize, gray] at (axis cs:66,3383) {Slot $n{+}5$};

\end{axis}

\node[draw, fill=white, inner sep=5pt, font=\small] at (10.3,0.8) {
    \begin{tabular}{@{}l@{}}
        \tikz{\fill[green!25, opacity=0.8] (0,0) rectangle (0.4,0.25);} Buy DEX, Sell CEX \\[2pt]
        \tikz{\fill[orange!25, opacity=0.8] (0,0) rectangle (0.4,0.25);} Buy CEX, Sell DEX
    \end{tabular}
};

\end{tikzpicture}

\caption{\centering The CEX-DEX arbitrage strategy over six Ethereum slots. The CEX mid price (blue line) updates continuously at millisecond frequency, while the DEX price (red line) updates only at 12-second slot boundaries following arbitrage execution. Green regions represent opportunities to buy on DEX and sell on CEX; orange regions represent the reverse direction.}
\label{fig:price_dynamics}
\end{figure}
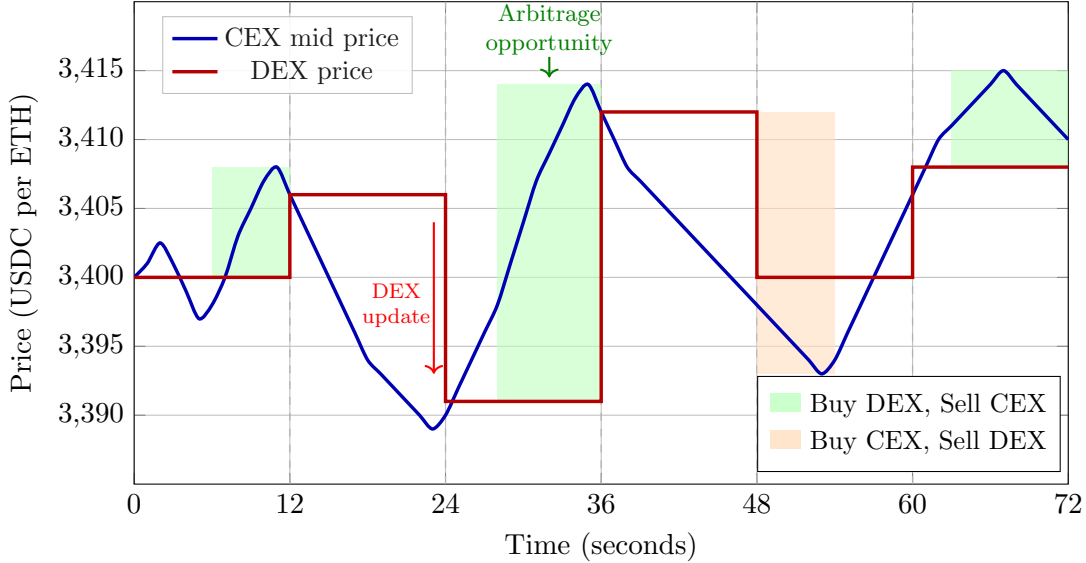

CEX-DEX arbitrage exploits price discrepancies between centralized and decentralized exchanges. When the DEX price is below the CEX bid, or above the CEX ask, an arbitrageur can buy on the DEX and sell on the CEX, or vice versa, to capture the spread. The strategy, which is depicted in \hyperref[fig:price_dynamics]{Figure 2}, appears straightforward, but the execution introduces significant complications.

The core risk is that the two legs of the trade are executed asynchronously. Assuming infinite liquidity and instant and guaranteed execution on Binance, the CEX leg settlement bears no risks, but the DEX leg needs to be included in a block. If the DEX transaction fails to land, due to competition from other arbitrageurs, network congestion, among other reasons, the trader is left with unhedged exposure. They have already committed to one side of the trade and must now close out the position at potentially unfavorable prices. 

We model an agent who explicitly accounts for this risk. The agent has a fixed probability $\alpha \in (0,1)$ of successfully landing the DEX leg of any attempted arbitrage. If the leg fails, the agent must decide how to manage the resulting exposure: close immediately on the CEX, retry the DEX trade in the next slot, or wait for better conditions. This fallback logic, combined with risk-averse preferences over profit variance, determines which opportunities the agent chooses to pursue.

Throughout the paper, we consider the arbitrage strategy of buying $Q$ units on the DEX and selling $Q$ units on the CEX \footnote{The opposite strategy of buying $Q$ units on the CEX and selling $Q$ units on the DEX has the same logic with signs flipped. When the DEX price is above the CEX ask, one would buy on CEX and sell on DEX. So, we consider the case mentioned above without loss of generality, as the other case is symmetric, and the logic remains the same.}. Let $p^{\text{CEX}}_{\text{bid}}(t)$ and $p^{\text{CEX}}_{\text{ask}}(t)$ denote the CEX bid and ask prices, and let $p^{\text{DEX}}(t_n)$ denote the DEX price at slot $n$, where $t_n = n\tau$ is the timestamp of slot $n$ and $\tau$ is the slot duration (i.e., 12 seconds by default, 1 second under faster execution). Key parameters include the time step $\delta$ (i.e., 1 second), the risk-aversion coefficient of the agent $\lambda$, and the minimum profit threshold $\theta$ for the entry.

\subsection{Simple Model}

In the simple model, the agent attempts to take every detected arbitrage opportunity without regard to execution risk. When an opportunity arises, which is defined as the CEX-DEX spread exceeding the pool fee, the agent immediately submits both legs of the trade. 

With probability $\alpha$, the DEX leg lands successfully, and the agent captures the full arbitrage profit:
\[\pi^{s} = Q \bigl(p^{\text{CEX}}_{\text{bid}} (t_n) - p^{\text{DEX}}(t_n) \bigr)\]

With probability $1 - \alpha$, the DEX leg fails. The agent has already sold on the CEX and must now close the short position. In the simple model, we assume that the agent simply buys back on the CEX one second later, paying the ask price and crossing the spread:
\[\pi^{f} = Q \bigl(p^{\text{CEX}}_{\text{bid}} (t_n) - p^{\text{CEX}}_{\text{ask}}(t_n + \delta) \bigr)\]

This model serves as a benchmark. In this setup, the agent's behavior is mechanical, as they act at every opportunity, so differences between 12-second and 1-second environments reflect only the change in opportunity frequency, not any strategic adaptation.

\subsection{Risk-Averse Model}
\label{sec:threetwo}

The risk-averse model endows the agent with two capabilities absent from the simple model: selective entry based on risk-adjusted expected profit and optimal fallback decision when trades fail. In this model, we study two main mechanisms: entry decision and fallback logic. \\

\noindent \textbf{Entry decision.} The agent enters an arbitrage opportunity only if the risk-adjusted expected profit exceeds a threshold $\theta$. Let $\pi(t_n)$ denote the random profit from attempting arbitrage at slot $n$. Hence, the expected profit is as follows:
\[\mathbb{E} \bigl(\pi (t_n) | \mathcal{F}_{t_n} \bigr) = \alpha Q \bigl(p^{\text{CEX}}_{\text{bid}} (t_n) - p^{\text{DEX}}(t_n) \bigr) + (1-\alpha) \mathbb{E} \bigl(V^f (t_{n,1}, 1)| \mathcal{F}_{t_n} \bigr) \]
\noindent where $V^f (t_{n,1}, 1)$ is the value function for the fallback problem, which will be defined below, and $\mathcal{F}_{t_n}$ is the information set at time $t_n$. Next, we define $X^s = Q \bigl(p^{\text{CEX}}_{\text{bid}} (t_n) - p^{\text{DEX}}(t_n) \bigr)$ as the profit conditional on success and $X^f = V^f (t_{n,1}, 1)$ as the random profit conditional on failure. Hence, the variance of profit is:
\[\text{Var} \bigl(\pi (t_n) | \mathcal{F}_{t_n} \bigr) = \alpha (1-\alpha) \biggl(X^s - \mathbb{E} \bigl(X^f | \mathcal{F}_{t_n} \bigr) \biggr)^2 + (1-\alpha) \text{Var} \bigl(X^f | \mathcal{F}_{t_n} \bigr)\]
Therefore, the agent enters if 
\begin{equation}
    \mathbb{E} \bigl(\pi (t_n) | \mathcal{F}_{t_n} \bigr) - \lambda \sqrt{\text{Var} \bigl(\pi (t_n) | \mathcal{F}_{t_n} \bigr)} \ge \theta
\end{equation}

\noindent \textbf{Fallback logic.} When the DEX leg fails, the agent holds an open CEX position that must eventually be closed. At each decision point $t_{n+k, m}$ (at slot $n+k$, subslot $m$) after $k$ failed DEX attempts, the agent chooses among three options:
\newline

\noindent \textit{1. Close immediately:} Buy back on the CEX at the current ask price. Hence, the profit is deterministic:
    \[\pi^{c} (t_{n+k, m}) = Q  \bigl(p^{\text{CEX}}_{\text{bid}} (t_n) - p^{\text{CEX}}_{\text{ask}}(t_{n+k, m}) \bigr)\]
    
\noindent \textit{2. Retry on DEX:} Submit a new DEX buy order at the current DEX price and wait for the next slot. If the retry succeeds (with probability $\alpha$):
\[\pi^{r} \bigl(t_{n+k, m} | L_{n+k} = 1\bigr) = Q  \bigl(p^{\text{CEX}}_{\text{bid}} (t_n) - p^{\text{DEX}}(t_{n+k}) \bigr)\]
\noindent On the contrary, if the retry fails (with probability $1-\alpha$), the agent enters the fallback problem again at the next slot with one additional failed attempt:
\[\pi^{r} \bigl(t_{n+k, m} | L_{n+k} = 0\bigr) = V^f (t_{n+k+1, 0}, k+1)\]
\noindent That is, the expected profit from retrying equals:
    \[\mathbb{E} \bigl(\pi^r (t_{n+k, m}) | \mathcal{F}_{t_{n+k, m}} \bigr) = \alpha Q  \bigl(p^{\text{CEX}}_{\text{bid}} (t_n) - p^{\text{DEX}}(t_{n+k}) \bigr) + (1-\alpha) \mathbb{E} \bigl(V^f (t_{n+k+1, 0}, k+1) | \mathcal{F}_{t_{n+k, m}} \bigr)\]
    
\noindent \textit{3. Wait:} Do nothing for one time step and reassess at $t_{n+k, m+1}$:
\[\pi^{w} (t_{n+k, m}) = V^f (t_{n+k, m+1}, k)\]

\noindent In the end, the agent selects the option with the highest risk-adjusted utility:
\[u^c (t_{n+k, m}) = \pi^{c} (t_{n+k, m})\]
\[u^r (t_{n+k, m}) = \mathbb{E} \bigl(\pi^{r} \bigl(t_{n+k, m}) | \mathcal{F}_{t_{n+k, m}}\bigr) - \lambda \sqrt{\text{Var} \bigl(\pi^r (t_{n+k, m}) | \mathcal{F}_{t_{n+k, m}} \bigr)}\]
\[u^w (t_{n+k, m}) = \mathbb{E} \bigl(\pi^{w} \bigl(t_{n+k, m}) | \mathcal{F}_{t_{n+k, m}}\bigr) - \lambda \sqrt{\text{Var} \bigl(\pi^w (t_{n+k, m}) | \mathcal{F}_{t_{n+k, m}} \bigr)}\]

\begin{figure}[htb!]
\centering
\tikzset{
    startstop/.style={
        rectangle, 
        rounded corners=10pt, 
        minimum width=4.2cm, 
        minimum height=1.5cm,
        text centered, 
        draw=black, 
        thick,
        text width=4cm,
        font=\footnotesize
    },
    process/.style={
        rectangle, 
        rounded corners=10pt,
        minimum width=4.2cm, 
        minimum height=1.5cm, 
        text centered, 
        draw=black, 
        thick,
        text width=4cm,
        font=\footnotesize
    },
    decision/.style={
        rectangle, 
        rounded corners=10pt,
        minimum width=4.5cm, 
        minimum height=2.8cm,
        text centered, 
        draw=black, 
        thick,
        text width=4.3cm,
        font=\footnotesize
    },
    arrow/.style={
        thick,
        -Stealth
    },
    label/.style={
        font=\scriptsize,
        text width=1.5cm,
        align=center
    }
}
\begin{tikzpicture}[node distance=2cm, auto, scale=0.85, every node/.style={transform shape}]
\node[font=\normalsize] at (0, 7.2) {Fallback Logic After Failed DEX Transaction};

\node (start) [startstop] at (0, 5.5) {
    \textbf{DEX Transaction Failed}\\[0.1cm]
    Open position with delta exposure\\[0.2cm]
    State: $(t_{n+k,m}, k)$ failed attempts
};

\node (evaluate) [process] at (0, 3) {
    \textbf{Evaluate Options}\\[0.1cm]
    $\max(U^c, U^r, U^w)$
};

\node (cex) [decision] at (-6, 0) {
    \textbf{Close via CEX}\\[0.1cm]
    - Buy $Q$ tokens on CEX now\\
    - Lock in loss/gain immediately\\
    - No variance (certain outcome)
};
\node (dex) [decision] at (0, 0) {
    \textbf{Retry via DEX}\\[0.1cm]
    - Submit new DEX order\\
    - Constraint: $m \leq M - \bar{M}$\\
    - Wait for the next confirmation step
};
\node (wait) [decision] at (6, 0) {
    \textbf{Wait for better conditions}\\[0.1cm]
    - Do nothing for $\delta$ seconds\\
    - Constraint: $m < M$\\
    - Reassess at $t_{n+k,m+1}$
};

\node (closed1) [process] at (-7, -3) {
    \textbf{Position Closed}\\[0.05cm]
    Final PnL realized
};
\node (closed2) [process] at (-2.25, -3) {
    \textbf{Position Closed}\\[0.05cm]
    Final PnL realized
};
\node (exposed) [process] at (2.25, -3) {
    \textbf{Still Exposed}\\[0.05cm]
    Enter fallback at\\
    $t_{n+k+1,0}$ (k+1)
};
\node (reevaluate) [process] at (7, -3) {
    \textbf{Re-evaluate}\\[0.05cm]
    Back to decision\\
    node at $t_{n+k,m+1}$
};
\draw [arrow] (start) -- (evaluate);
\draw [arrow] (evaluate) -- (cex);
\draw [arrow] (evaluate) -- (dex);
\draw [arrow] (evaluate) -- (wait);
\draw [arrow] (cex) -- (closed1);
\draw [arrow] (dex) -- node[label, left=0.3cm] {Success ($\alpha$)} (closed2);
\draw [arrow] (dex) -- node[label, right=0.3cm] {Failure ($1-\alpha$)} (exposed);
\draw [arrow] (wait) -- (reevaluate);
\end{tikzpicture}
\caption{Decision tree showing fallback logic after a failed DEX transaction.}
\label{fig:fallback_logic}
\end{figure}
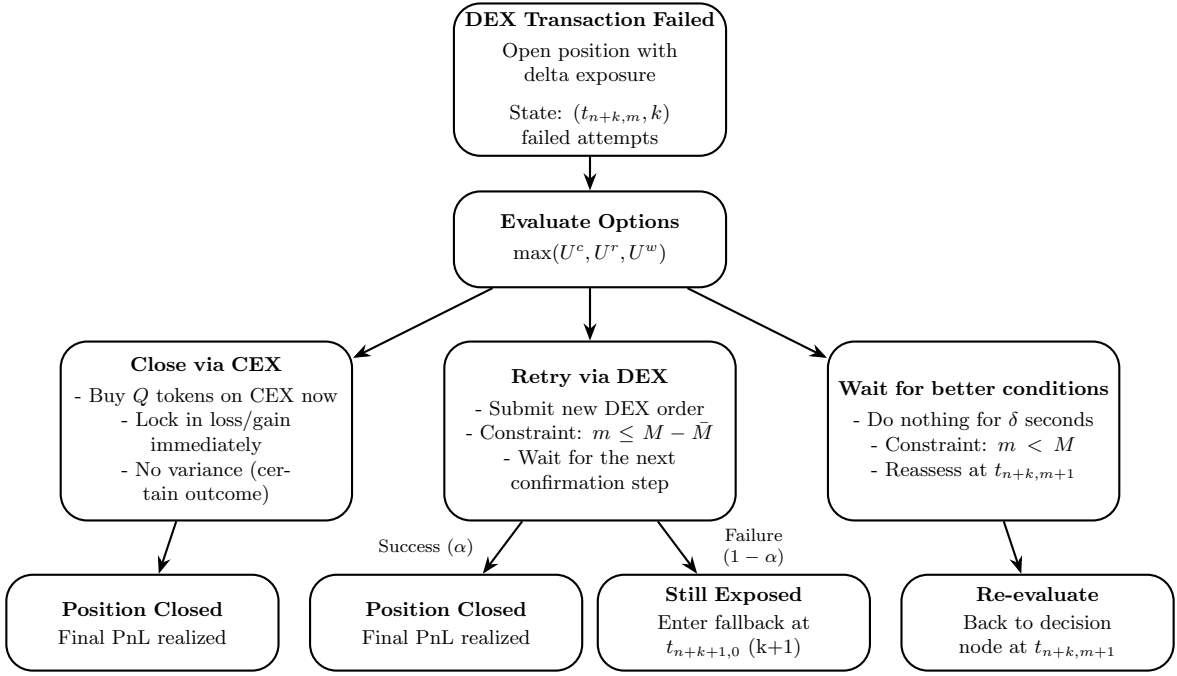

\noindent Hence, the value function is as follows:
\begin{equation}
V^f (t_{n+k, m}, k) = \max \bigl\{u^c (t_{n+k, m}), u^r (t_{n+k, m}), u^w (t_{n+k, m}) \bigr\}
\end{equation}
\noindent subject to constraints: retry is only available if there is sufficient time before slot end ($m \le M - \bar{M}$), wait is only available within a slot ($m < M$), and after $\bar{k}$ failed attempts, the agent must close. The fallback logic is depicted in \hyperref[fig:fallback_logic]{Figure 3}. \\

\noindent \textbf{Price dynamics for expectation computation.} Computing the value function requires modeling future prices. Here, we assume CEX log-returns follow a Gaussian random walk:
\[\log p^{\text{CEX}}_{\text{mid}}(t_{n+k,m}) = \log p^{\text{CEX}}_{\text{mid}}(t_{n+k,0}) + \sum^m_{j=1} \varepsilon_j, \quad \varepsilon_j \sim \mathcal{N}(0, \sigma^2 \delta)\]
\noindent with bid and ask prices given by $p^{\text{CEX}}_{\text{bid}} = (1-\beta)p^{\text{CEX}}_{\text{mid}}$ and $p^{\text{CEX}}_{\text{ask}} = (1+\beta)p^{\text{CEX}}_{\text{mid}}$ where $\beta$ is the half-spread. DEX prices are updated only at slot boundaries and track the CEX mid-price with mean-reverting basis risk $\eta$:
\[p^{\text{DEX}} (t_{n+k}) = p^{\text{CEX}}_{\text{mid}} (t_{n+k}) + \eta_{n+k}\]

\noindent \textbf{Computational approach.} We solve the value function by backward induction with Monte Carlo estimation at each node. The terminal condition is forced closure at $k = \bar{k}$:
\[V^f (t_{n+\bar{k}, m}, \bar{k}) = Q  \bigl(p^{\text{CEX}}_{\text{bid}} (t_n) - p^{\text{CEX}}_{\text{ask}}(t_{n+\bar{k}, m}) \bigr)\]

\noindent Recursion proceeds in two nested loops: an outer loop over failed attempts ($k = \bar{k} - 1, ..., 0$) and an inner loop over subslots within each slot ($m = M, ..., 0$). At each state, we simulate $N$ price paths to estimate the expectation and variance of each option, and then select the option with the highest risk-adjusted utility.

\section{Simulation Design}

The simulation models the competitive dynamics between agents operating CEX-DEX arbitrage bots. Agent 1 represents the participant whose behavior we analyze in detail, while Agent 2 represents the aggregate behavior of all other market participants. The two agents compete for arbitrage opportunities, but only one can successfully execute the DEX leg of any given trade. 

When an arbitrage opportunity arises, multiple agents might attempt to capture it, but blockchain constraints ensure only one transaction can succeed. When Agent 1 attempts an opportunity, they win with probability $\alpha$. When Agent 1 fails or chooses not to attempt to land their DEX transaction, Agent 2 automatically succeeds and captures the arbitrage profit. Conversely, when Agent 1 succeeds, Agent 2's transaction fails. 

This competitive structure introduces execution risk that fundamentally shapes agent behavior. In the simple model, both agents attempt every detected opportunity regardless of expected outcomes. In the risk-averse model, Agent 1 evaluates each opportunity's risk-adjusted expected value before entering, while Agent 2 continues to pursue all opportunities. When Agent 1 declines to enter based on unfavorable risk-return characteristics, Agent 2 captures the arbitrage. Agent 2 represents a simplified aggregate of market participants who close failed positions immediately on the CEX, while Agent 1 employs the risk-averse fallback logic described in \hyperref[sec:threetwo]{Section 3.2}, potentially waiting for better conditions or retrying on the DEX. 

The core objective of the simulation is to quantify how faster execution guarantees change arbitrageur behavior and market outcomes. For each of the four configurations listed below, we run parallel simulations under two execution regimes. In the 12-second execution regime, DEX transactions can only be executed at Ethereum slot boundaries, occurring every 12 seconds: this represents the simple environment where arbitrageurs face substantial execution windows and corresponding uncertainty. In the 1-second execution regime, DEX transactions can be executed every second through subslot confirmations: this represents the improved environment where arbitrageurs receive faster execution guarantees. \\

\noindent \textbf{Configuration Space.} We evaluate agent behavior across multiple environmental configurations to assess the robustness of our findings. Each configuration combines two binary design choices that capture different aspects of market microstructure.

The first choice is CEX-DEX price reversion. When enabled, subslot DEX prices gradually adjust toward CEX prices between arbitrage events using the regression-based mechanism described in \hyperref[sec:twotwo]{Section 2.2}. This captures informed trading that occurs independently of direct arbitrage.

The second choice is noise trading. When enabled, the simulation incorporates random non-arbitrage transactions whose frequency and price impact are sampled from empirical distributions estimated from data. This captures retail flow and other trading activity unrelated to CEX-DEX arbitrage.

These two binary choices yield four distinct configurations, ranging from a minimal environment with no reversion and no noise, to an environment incorporating both components. The results of these configurations are then compared to an agent with the current 12-second confirmation interval. \\ 

\noindent \textbf{Key assumptions.} The simulation relies on several key assumptions. On the execution side: all validators are opted in to providing fast execution guarantees; hence, there are no missed slots, and DEX transactions can occur every second; gas fees are zero; agents cannot run out of capital, meaning liquidity constraints do not affect execution, and there are no fees other than DEX pool fees. On the arbitrage side: every top-of-block arbitrage opportunity is executed; arbitrages are executed with the optimal trade size, moving the DEX price to one pool fee from CEX bid or ask; infinite liquidity exists at the best bid and ask on Binance, and execution on Binance is instant and guaranteed. On market structure: DEX pools are constant-product (Uniswap v2-style); liquidity is constant except for the increase from swap fees, and all pools start with identical initial liquidity for comparability. \\ 

\noindent \textbf{Parameters.} For parameters used by both models, Agent 1's winning probability is set to $\alpha = 0.35$, and the CEX-DEX reversion interval window is 300 seconds. For parameters specific to the risk-averse model, the risk-aversion coefficient is $\lambda = 0.01$, the entry threshold is $\theta = 0$, and Monte Carlo estimation uses 16 paths at each node. The decision wait horizon is 3 seconds, and the maximum wait time is also 3 seconds. Robustness tests vary $\alpha \in \{0.20, 0.35, 0.50\}$ and $\lambda \in \{0, 0.01, 0.03\}$. \\

\noindent \textbf{Combined effect of changes.} To assess the overall impact across all three pools, we compute a weighted average of the pool-specific changes in transaction counts and volume. The weights reflect each pool's historical share of arbitrage activity during the observation period.
For transaction counts, the 30, 5, and 1 basis point pools represent 3.7\%, 33.1\%, and 63.2\% of total arbitrage transactions, respectively. For trading volume, these pools account for 28.6\%, 50.3\%, and 21.1\% of total arbitrage volume, respectively.

\section{Results}

We examine the impact of reducing the confirmation interval from 12 seconds to 1 second on transaction frequency and trading volume across different configurations and fee-tier pools, comparing simple and risk-averse agent models. All reported ranges reflect aggregated results across all fee levels and configurations for an agent with a 35\% probability of successfully executing a DEX leg. \\ 

\noindent \textbf{Transaction Frequency.} We measure transaction frequency as the number of transactions landed by the agent. In the simple model, transaction counts increase substantially across all pools. Across all configurations and fee tiers, the simple model exhibits increases ranging from 218\% to 663\%. 

The risk-averse model exhibited an even greater increase in transaction frequency, with a transaction count increase ranging from 294\% to 1386\%. However, it is important to note that most scenarios fall below an increase of 600\% with the 1386\% being present in a single configuration where, while the percentage change is big, the absolute number of transactions is quite small.

The pattern across fee tiers reflects the economics of arbitrage at different spread levels. Lower-fee pools have tighter spreads and more marginal opportunities. Under a 12-second confirmation, many of these opportunities are not worth pursuing because the execution window is long enough that adverse price moves can wipe out the slim profit margin. Under 1-second confirmation, the risk is compressed, and more marginal opportunities become viable.

The weighted average of transaction number increase across different pools varies between 356\% and 412\% for the simple model depending on the configuration, with the average of noise with and without reversion configurations being 371\%. Similarly, for the risk-averse model, the weighted increase varies between 378\% and 567\%, with the average of two configurations being 535\%. \\

\noindent \textbf{Trading Volume.} Trading volume, measured in terms of Ethereum, shows more nuanced patterns than changes in number of transactions. In the simple model, the volume changes exhibit increases ranging from 98\% to 273\% depending on the configuration. Similarly to the trading frequency, the increase is closer to the lower end of the range for the 30 basis point pool and closer to the higher end for the 5 and 1 basis point pools. The risk-averse model showed similar patterns with increases ranging from 121\% to 375\%. 

The volume increases are concentrated in lower-fee pools. The intuition here is as follows: when more marginal opportunities become viable, the average trade size may fall (as marginal opportunities tend to have smaller optimal trade sizes), but the total number of trades increases by enough to raise aggregate volume. In higher-fee pools, where opportunities are already large and infrequent, faster confirmations do not as dramatically expand the viable opportunity set. 

It is important to note that these ranges represent outcomes under different configurations; however, since gas fees are not incorporated into the model, the upper bounds of these ranges might overestimate the increase for the 1-basis-point and 5-basis-point pools, where the economic viability of frequent small transactions would be constrained by transaction costs in practice. 

When weighing the volume increases based on the arbitrage volume across different pools, the average increase in volume for the three pools varies between 148\% and 211\% for the simple model, with the average being 168\%. Similarly, for the risk-averse model, the overall increase varies between 159\% and 243\%, with the average being 203\%. \\

\noindent \textbf{Robustness Tests.} To assess the robustness of these findings, we conducted an additional set of simulations for the risk-averse model under the reversion and noise trading configuration. We varied Agent 1’s win probability $\alpha \in \{0.20, 0.35, 0.50\}$ and the risk aversion parameter $\lambda \in \{0, 0.01, 0.03\}$, and compared outcomes when the confirmation interval is reduced from 12 seconds to 1 second. 

In the 5 basis point pool, across all combinations in this grid, transaction counts increased by 416–530\%, and ETH trading volume increased by 146–186\%. In the 30 basis point pool, transaction counts increased by 1152–1468\%, and ETH volume changes ranged between a 317\% increase and a 385\% increase. Overall, changes in transaction counts and volume are of the same magnitude when comparing agents with identical parameters under 12-second versus 1-second confirmation intervals, suggesting that the main results are robust to reasonable variations in win probability and risk aversion.

\newpage
\section{Conclusion}

This paper examines how reducing blockchain execution times affects decentralized exchange activity, focusing on the behavior of CEX-DEX arbitrageurs. Our contribution is as much methodological as empirical: we develop a simulation framework combining empirical price anchoring, noise-trading dynamics, CEX–DEX arbitrage mechanics, and a risk-averse decision model. The agent we model takes execution risk seriously, incorporating DEX transaction uncertainty into both entry decisions and fallback strategies. This setup captures a fundamental reality of on-chain trading that prior work has largely abstracted away.

Under reasonable assumptions about agent risk preferences and inclusion probabilities, the shift from a 12-second to a 1-second execution environment increases the number of arbitrage transactions by 535\%, with effects concentrated in lower-fee pools where marginal opportunities become newly viable. Faster execution also reduces profit variance by compressing the window for adverse price moves during failed-trade recovery. 

Several limitations are worth mentioning. Our model assumes zero gas fees, which may overstate the viability of small trades in low-fee pools. Also, the agent's probability of landing DEX trades is fixed rather than endogenously determined by competition; in equilibrium, faster execution might attract more arbitrageurs and compress $\alpha$. These extensions remain for future work.

\newpage

\appendix
\section{Appendix}

Below are tables with exhaustive results for different configurations.

\subsection*{Table A1}
Changes in the simple agent's metrics for the 30 bps pool when confirmation frequency is decreased to 1 second.
\begin{table}[H]
\centering
\begin{tabular}{lcccc}
\toprule
Configuration & $\Delta$ PnL & $\Delta$ ETH Vol. & $\Delta$ USDC Vol. & $\Delta$ Txns \\
\midrule
no reversion, no noise & $+113\%$ & $+118\%$ & $+116\%$ & $+294\%$ \\
no reversion, noise & $+97\%$ & $+98\%$ & $+97\%$ & $+218\%$ \\
reversion, no noise & $+276\%$ & $+273\%$ & $+265\%$ & $+663\%$ \\
reversion, noise & $+218\%$ & $+211\%$ & $+205\%$ & $+478\%$ \\
\bottomrule
\end{tabular}
\end{table}

\subsection*{Table A2}
Changes in the simple agent's metrics for the 5 bps pool when confirmation frequency is decreased to 1 second.
\begin{table}[H]
\centering
\begin{tabular}{lcccc}
\toprule
Configuration & $\Delta$ PnL & $\Delta$ ETH Vol. & $\Delta$ USDC Vol. & $\Delta$ Txns \\
\midrule
no reversion, no noise & $+138\%$ & $+158\%$ & $+157\%$ & $+308\%$ \\
no reversion, noise & $+135\%$ & $+151\%$ & $+150\%$ & $+274\%$ \\
reversion, no noise & $+147\%$ & $+174\%$ & $+172\%$ & $+345\%$ \\
reversion, noise & $+144\%$ & $+165\%$ & $+164\%$ & $+313\%$ \\
\bottomrule
\end{tabular}
\end{table}

\subsection*{Table A3}
Changes in the simple agent's metrics for the 1 bp pool when confirmation frequency is decreased to 1 second.
\begin{table}[H]
\centering
\begin{tabular}{lcccc}
\toprule
Configuration & $\Delta$ PnL & $\Delta$ ETH Vol. & $\Delta$ USDC Vol. & $\Delta$ Txns \\
\midrule
no reversion, no noise & $+195\%$ & $+203\%$ & $+202\%$ & $+420\%$ \\
no reversion, noise & $+207\%$ & $+200\%$ & $+199\%$ & $+408\%$ \\
reversion, no noise & $+207\%$ & $+205\%$ & $+204\%$ & $+432\%$ \\
reversion, noise & $+212\%$ & $+202\%$ & $+201\%$ & $+420\%$ \\
\bottomrule
\end{tabular}
\end{table}

\subsection*{Table A4}
Changes in the risk-averse agent's metrics for the 30 bps pool when confirmation frequency is decreased to 1 second.
\textbf{Note:} For the table below, it is important to note that while the changes in the number of transactions in terms of percentages are great for certain configurations, the numbers are small in absolute terms.
\begin{table}[H]
\centering
\begin{tabular}{lcccc}
\toprule
Configuration & $\Delta$ PnL & $\Delta$ ETH Vol. & $\Delta$ USDC Vol. & $\Delta$ Txns \\
\midrule
no reversion, no noise & $+114\%$ & $+121\%$ & $+119\%$ & $+294\%$ \\
no reversion, noise & $+119\%$ & $+126\%$ & $+124\%$ & $+336\%$ \\
reversion, no noise & $+282\%$ & $+274\%$ & $+267\%$ & $+639\%$ \\
reversion, noise & $+365\%$ & $+375\%$ & $+365\%$ & $+1386\%$ \\
\bottomrule
\end{tabular}
\end{table}

\subsection*{Table A5}
Changes in the risk-averse agent's metrics for the 5 bps pool when confirmation frequency is decreased to 1 second.
\begin{table}[H]
\centering
\begin{tabular}{lcccc}
\toprule
Configuration & $\Delta$ PnL & $\Delta$ ETH Vol. & $\Delta$ USDC Vol. & $\Delta$ Txns \\
\midrule
no reversion, no noise & $+135\%$ & $+158\%$ & $+157\%$ & $+307\%$ \\
no reversion, noise & $+137\%$ & $+162\%$ & $+161\%$ & $+444\%$ \\
reversion, no noise & $+145\%$ & $+174\%$ & $+173\%$ & $+345\%$ \\
reversion, noise & $+147\%$ & $+179\%$ & $+178\%$ & $+500\%$ \\
\bottomrule
\end{tabular}
\end{table}

\subsection*{Table A6}
Changes in the risk-averse agent's metrics for the 1 bp pool when confirmation frequency is decreased to 1 second.
\begin{table}[H]
\centering
\begin{tabular}{lcccc}
\toprule
Configuration & $\Delta$ PnL & $\Delta$ ETH Vol. & $\Delta$ USDC Vol. & $\Delta$ Txns \\
\midrule
no reversion, no noise & $+151\%$ & $+205\%$ & $+204\%$ & $+419\%$ \\
no reversion, noise & $+151\%$ & $+205\%$ & $+205\%$ & $+544\%$ \\
reversion, no noise & $+158\%$ & $+208\%$ & $+207\%$ & $+472\%$ \\
reversion, noise & $+161\%$ & $+206\%$ & $+206\%$ & $+554\%$ \\
\bottomrule
\end{tabular}
\end{table}

\subsection*{Table A7}
Combined effect across three pools, weighted by historical transaction counts and volume, showing changes when confirmation frequency is decreased to 1 second.
\begin{table}[H]
\centering
\begin{tabular}{lcccc}
\toprule
 & \multicolumn{2}{c}{Simple} & \multicolumn{2}{c}{Risk-averse} \\
\cmidrule(lr){2-3} \cmidrule(lr){4-5}
Configuration & $\Delta$ ETH Vol. & $\Delta$ Txns & $\Delta$ ETH Vol. & $\Delta$ Txns \\
\midrule
no reversion, no noise & $+158\%$ & $+378\%$ & $+159\%$ & $+378\%$ \\
no reversion, noise & $+148\%$ & $+356\%$ & $+163\%$ & $+503\%$ \\
reversion, no noise & $+211\%$ & $+412\%$ & $+212\%$ & $+437\%$ \\
reversion, noise & $+188\%$ & $+387\%$ & $+243\%$ & $+567\%$ \\
\bottomrule
\end{tabular}
\end{table}

\end{document}